# ANALYSIS OF DIGITAL KNAPSACK BASED SEALED BID AUCTION

Navya Chodisetti

Abstract: The need of totally secure online auction has led to the invention of many auction protocols. But as new attacks are developed, auction protocols also require corresponding strengthening. We analyze the auction protocol based on the well-known mathematical public-key knapsack problem for the design of asymmetric public-key knapsack trapdoor cryptosystem. Even though the knapsack system is not cryptographically secure, it can be used in certain auction situations. We describe the limitations of the protocol like detecting and solving the tie between bidders, malicious behavior of participants and also selection of price set by the seller and offer solutions.

## INTRODUCTION

The word "auction" is derived from the Latin word called *auctus*, which means "increase" [from a small to a greater price]. The normal design of auction procedure is the ascending auction where the price of the product is raised such that only one bidder remains in the game. But the term "auction" now covers a wide variety of forms. Some auctions are descending to the extent that the bidder stops the process, some are public and observable in open auction and some are private in case of written and sealed bid auctions. The need of totally secure online auction has led to the invention of many new auction protocols. The development of these auctions is based on combining old protocols with the new science of public-key cryptography, complexity theory [1]-[8], and quantum information processing [9],[10].

Traditional auctions are analyzed into four basic types which are often referred as standard auctions: these are the English auction, the Dutch auction, the first-price sealed-bid auction, and the second price auction.

**English auction.** The English auction is also known as the open, oral, or ascending-bid auction. This is the oldest as well as the most wide spread auction mechanism. In this auction the auctioneer starts the auction by announcing the starting bid. Then the participant's bids increases the price successively, which must be higher than the current price to get accepted. The auction runs until one bidder remains in process. This auction can be run either by auctioneer announcing the prices or by bidders submitting bids shouting out loud or electronically in case of online auctions.

The price paid by the winner in an English auction depends on the price at which her/her rival exits the auction. If the bidder quits when the price reaches his/her valuation, then the price paid by the winner is equal to the second-highest valuation among bidders' valuations.

**Dutch auction.** The Dutch auction is an open descending-price counterpart of English auction. In this auction, the auctioneer starts at a very high price, so high that no bidder is interested in buying the object at this price. Then the price is gradually lowered, until one bidder signals his acceptance to the current price. Dutch auctions do not progressively reveal their valuation for the auctioned item, since only information that bidders have is that the auction in over. Each bidder must choose how high to bid without the knowledge of other bidders' interest in the item. The Dutch auction only reveals the price at which the winner values the good for sale.

To consider an example of Dutch auction, consider a software company that starts the auction at $3,000 for its shares. If there are no bidders, the price is lowered by $200. The shares will be sold once the bidder accepts the last announced price by the auctioneer, say $1,800. Once all bids are placed, the bids are considered from highest bid down until all of the shares are assigned. The price that each bidder pays is based on the lowest price of all allotted bidders. Therefore, even if A bid $300 for his 2,000 shares, if the last successful bid is $100, A will have to pay $100 for his 2,000 shares.

**First-price sealed-bid auction (FPSB).** In this auction, bidders simultaneously submit a single bid to the auctioneer in a sealed envelope without knowing other bidders' bids. By the end of the deadline, the envelopes are opened and submitted bids are evaluated. The bidder with the highest bid is declared as the winner. There is no interaction among bidders as in English auction and the bidders submit their bid based on their willingness to pay and assuming others bidders' valuation.

**Second-price sealed-bid (SPSB).** This is similar to the first-price sealed bid. Here each bidder privately submits a single bid and the winner is the bidder who makes the highest bid.

Example: Let the highest bid other than me be denoted by R. Assuming that my value for the product is $2000. When R < 2000 any bid above R will make me win the bid and any bid below R will make me lose the bid and when R > 2000 bidding above R will me make me lose money but win the object and bidding below R gives me 0 payoff.

**Digital Auctions.** Auctions have become a major phenomenon in the field of electronic commerce. A digital auction is viewed as a set of electronic protocols which allows a collection of bidders to buy a thing at an auction with the lowest possible price, while a seller wants the bidder to buy the product with highest possible price.

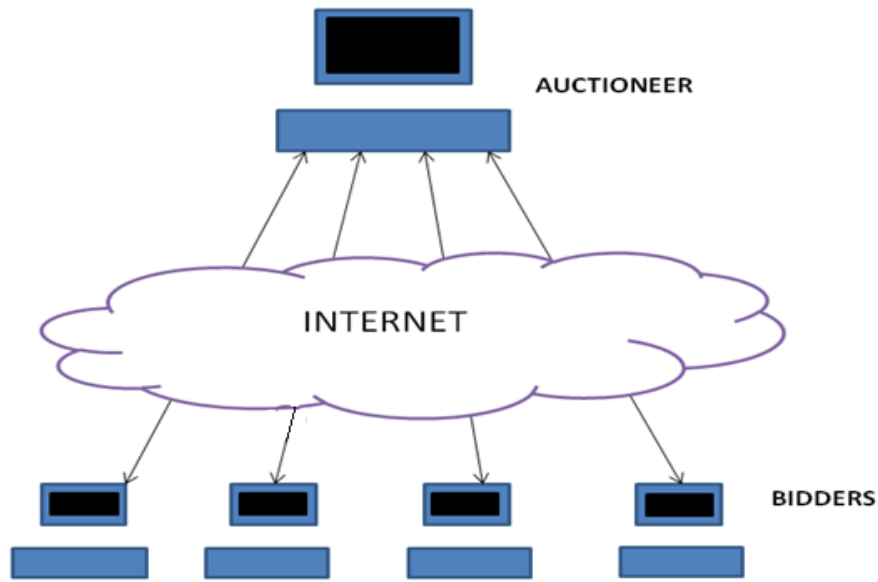

Figure 1 : Digital Auction

They are many factors concerning the design of a successful online auction protocol that include the following:

- Secrecy of bidding price: Bidding prices of bidders should not be revealed even after the completion of the bid.
- Validity of a successful bid: Anyone participating in the bid should be able to verify the validity of the bid.
- Fairness: only the registered bidder must be able to bid the auction.

- Anonymity of winner: No information should be revealed about the identity of any bidder.
- Correctness: Determining the correct winner of the bid and its corresponding selling price.

To satisfy the above requirements, the use of cryptographic techniques becomes essential [11]-[22].

The online auction site eBay was established in 1995, and it has now expanded into many countries and it makes billions of dollars each year. Many believe that in spite of its success the auction mechanism used by eBay and other auction sites are not perfect. For example, eBay runs many simultaneous sequential auctions and the same kind of goods may be on going in hundreds, sometimes thousands of auctions, which makes it difficult for a potential buyer to choose an auction to place his/her bid. Another problem with these auctions is that eBay typically finishes its auction at a fixed time, which allows the bidders to place a bid moments before the auction terminates causing loss of revenue for both seller and eBay.

**OVERVIEW OF BASIC TOOLS**
**Oblivious Transfer.** Oblivious transfer is a cryptographic protocol where the sender transfers one of the potentially many pieces of information to a receiver, but remains oblivious as to which piece has been transferred. The idea of oblivious transfer (OT) was first proposed by Rabin [23].

In OT the sender S has only one secret $m$ and this message have a probability 0.5 to reach the receiver R. The receiver $R$ does not want $S$ to know, if it gets $m$ or not. For $OT_2^1$, sender S has two secret messages $m_1$ and $m_2$, the receiver will get one of them by the choice of receivers choice. The bit chosen by the receiver is unknown to sender because he/she wants it that way. And also receiver must not know any information other than what he/she has chosen. Similarly $OT_Z^1$ is a natural extension of $OT_2^1$ in case of z secrets. Where the secrets $m_1, m_2, m_3 \ldots\ldots m_z$ and is willing to disclose only one among them to receiver. Oblivious transfer is the fundamental primitive in many cryptographic applications and secure distributed computations and has many applications such as private information retrieval (PIR), fair electronic contract signing [24].

OT can be used to generate random numbers, but the randomness can only be guaranteed on a computational complexity basis. Randomness can also be viewed probabilistically [25] or from mathematical and physical basis [26]-[29].

**Knapsack Problem.** The knapsack problem is a mathematically attractive proposition for cryptography and the Merkle and Hellman public key asymmetric-key cryptography was based on knapsack problem [30]. Although it was eventually broken [2], it still is a useful scheme to use in certain applications.

Assume key $K = (k_1, k_2 ....... k_n)$ where $k_n$'s are integers and n is the plain text bit length. Let $S = (s_1, ........ s_n)$ be a plain text where, $s_n \in \{0,1\} \forall n = (1.....n)$. then the encryption of the plain text S using the knapsack cryptosystem is done by the formula

$$Z = K.S = \sum_{i=1}^{n} k_i x_i .$$

From the formula calculating Z is really simple whereas recovery of S from the same is really difficult since K is randomly chosen. But the knapsack problem becomes really simple if we choose the random numbers in a way other than choosing each element of K is larger than the sum of the preceding elements.

If $k_i > \sum_{j=1}^{i-1} k_j \forall i = (1....n)$ the cipher can be generated as $Z = \sum_{j=1}^{n} k_j x_j$.

The plain text X can be recovered from the key K and the cipher text $z_n$.

**KNAPSACK BASED AUCTION PROTOCOL**
This survey examines the auction protocol described by Ibrahim [20] where there is a seller and a set of n bidders come to an agreement on the selling price of a certain product. The role of the auctioneer becomes obsolete in this case since only the seller and the set of bidders are involved in this protocol.

Firstly the seller publicly announces a set of prices in an increasing order to the bidders. Bidders are allowed to choose any price from the set of prices and make a sealed bid. Once the bidder makes the bid the problem related to the security of the bid must be considered. To do this seller defines a set of secret keys, each key corresponding to price in price set in a super increasing order. Once the bidder makes his/her choice by selecting a price from the price set the seller assigns him/her with a secret key; this is later used by bidder to perform the additive sharing with other bidders to stay anonymous.

**Seller Initialization the Following:**
1. Select a set of prices $P = \{p_1, p_2......p_n\}$ where $p_j < p_i \forall j < i$ in an increasing order.
2. Once the prices set P is selected now the seller defines a set of n super increasing random integers, $C = \{c_1, c_2......c_n\}$ where $c_i > \sum_{j=1}^{i-1} c_j \forall i = (1.....n)$, such that the seller assigns each $c_i$ to respective $p_i$
3. There is an another set where the seller select a set of random integers $R = \{r_1, r_2......r_i\}$ such that $\sum_{n=1}^{i} r_n = 0 \bmod q \wedge r_n \in N_p \forall n = (1....i)$ where q is a large prime $q > \sum_{j=1}^{k} c_j$
4. The seller defines vector of k flags $F = \{f_1, f_2......f_k\}$ where $f_i = (1......k) = 0 \forall i$ where $f_i \in \{0,1\}$.

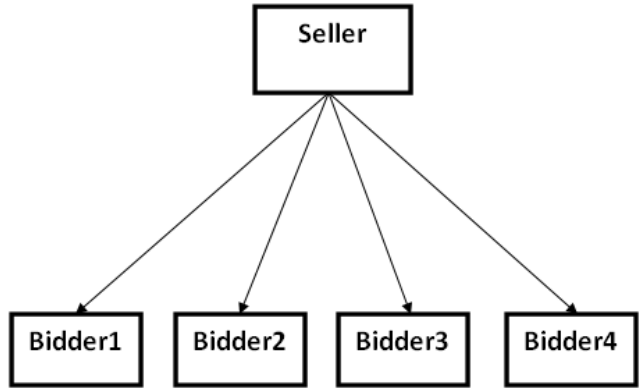

Figure 2: Seller and bidder in Auction

**Bidders Makes the Following Choices:**
1. Bidder $B_i \forall i(1....n)$ selects a price from set P.

**Oblivious Transfer of Strings to Securely Transfer the Secret Code:**

1. Bidder $B_i$ interacts with the seller through Oblivious transfer $OT_k^1$ of strings to get the secret randomized code, $c_{xj}^1 = c_{xj} + r_j \mod q$ by choosing a respective secret code from set C.

**Bidders Share the Secret by Additive Sharing:**
1. Each bidder $B_i$ divides his secret code into i random values, such that $c_{ij}^1 = \sum_{v=1}^{i} m_{j,v}$.
2. Then each bidder $B_i$ privately sends $m_{j,v}$ to every other bidder $B_v \forall v = (1...i)$
3. Bidder then sums all the bidder shares to compute the additive share $\sigma_j = \sum_{v=1}^{i} m_{jv}$
4. Then these values are collected from bidder to compute $\sigma_k$.

**Seller Solves the Knapsack Problem:**
1. Seller computes the knapsack value, $\sigma_k = \sum_{n=1}^{i} \sigma_n \mod q$
2. If $\sigma_k < c_k$ then set $f_k = 0$ and $\sigma_{k-1} = \sigma_k$, then set $f_k = 1$ and $\sigma_{k-1} = \sigma_k - c_k$
3. States of the flags in F are determined by solving the knapsack problem.

**Final Announcement of Winner:**
1. The seller picks the highest price indicated by the vector $F$ which contains the flags corresponding to the prices $p$.
2. The seller requests the winning bidder to identify himself.
3. Bidder proves his case to seller by displaying the secret code.

**Example 1:** This example has a price set of 9 bidding values with 4 bidders participating in the auction.
- Firstly consider that the seller offering a product for the following 9 bidding prices $p$= {10, 100, 200, 250, 300, 350, 400, 450, 500} all in dollars.
- Now the seller also chooses a set C with 9 super increasing random integers
  C= {5, 9, 15, 30, 60, 120, 250, 500, 1000}
- Pick a prime number q in such a way that q>1984 which we considered to be 1987 in this example.
- Consider that they are 4 bidders participating in this bid, the seller chooses 4 random and independent integers and sets the vector $R = \{900, 700, 300, 87\}$ satisfying $900+700+300+87 \equiv 0 \mod 1987$.

- Suppose there are 4 bidders $B = \{B_1, B_2, B_3, B_4\}$ willing to bid.
- Assume that the following bids are made $B_1 \leftarrow 10\$, B_2 \leftarrow 200\$, B_3 \leftarrow 450\$, B_4 \leftarrow 350\$$

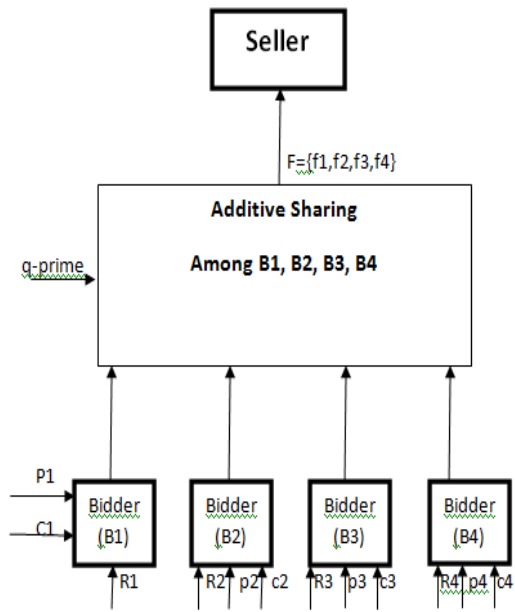

Figure 3 : Auction Protocol

- Now the seller interacts with each bidder through oblivious transfer of strings to securely transfer the randomized secret code corresponding to the bidders selected price index.

  $B_1 \leftarrow 5+900\text{mod}1987=905$

  $B_2 \leftarrow 15+700\text{mod}1987=715$

  $B_3 \leftarrow 500+300\text{mod}1987=800$

  $B_4 \leftarrow 120+87\text{mod}1987=207$

  Now bidder divides his secret code into 4 additive shares

  $B_1$ : 200+200+200+305=905

  $B_2$ : 100+115+400+100=715



$B_3$: 150+250+100+300=800

$B_4$: 50+70+80+7=207

Now each bidder computes his additive share

$B_1$: $\sigma_1$=200+100+150+50=500

$B_2$: $\sigma_2$ =200+115+250+70=635

$B_3$: $\sigma_3$=200+400+100+80=780

$B_4$: $\sigma_4$=305+100+300+7=712

Each bidder sends his additive share to the seller and then the seller computes his knapsack problem

$\sigma_8$=500+635+780+712=2627mod1987=640

Now the seller starts solving the knapsack problem and sets of the flag.

$\sigma_9 < c_9 \Rightarrow f_9 = 0, \sigma_8 = 640;$

$\sigma_8 > c_8 \Rightarrow f_8 = 1, \sigma_8 - c_8 = 640 - 500 = 140;$

$\sigma_7 < c_7 \Rightarrow f_7 = 0, \sigma_6 = 140;$

$\sigma_6 > c_6 \Rightarrow f_6 = 1, \sigma_6 - c_6 = 140 - 120 = 20;$

$\sigma_5 < (c_5, c_4) \Rightarrow f_5, f_4 = 0, \sigma_3 = 20;$

$\sigma_3 > c_3 \Rightarrow f_3 = 1, \sigma_3 - c_3 = 20 - 15 = 5$

$\sigma_2 < (c_2) \Rightarrow f_2 = 0, \sigma_2 = 5;$

$\sigma_1 > c_1 \Rightarrow f_1 = 1$

Final set f={1,0,1,0,0,1,0,1,0} and seller announces that the highest price as 450$. The winning bidder announces himself by providing his secret 800 to the seller. And second highest price can also be announced respectively using the same process.

**Example 2:** This example has a price set of ten bidding values with 3 bidders participating in the bid.
- For our second example consider that the seller offering a product for the following 10 bidding prices $p$= {100, 200, 300, 400, 500, 600, 700, 800, 900, 1000} all in dollars.
- Now the seller also chooses a set C with 10 super increasing random integers
  C= {3, 7, 15, 30, 65, 140, 320, 650, 1300, 2600}
- Pick a prime number $q$ in such a way that $q$>5130 which we considered to be



5209 in this example.

- Consider that they are 3 bidders participating in this bid, the seller chooses 4 random and independent integers and sets the vector $R = \{2500, 1500, 1209\}$ satisfying $2500+1500+1209 \equiv 0 \mod 5209$.
- Suppose there are 3 bidders $B = \{B_1, B_2, B_3\}$ willing to bid.
- Assume that the following bids are made $B_1 \leftarrow 200\$, B_2 \leftarrow 400\$, B_3 \leftarrow 1000\$$
- Now the seller interacts with each bidder through oblivious transfer of strings to securely transfer the randomized secret code corresponding to the bidders selected price index.

  $B_1 \leftarrow 7+2500 \mod 5209 = 2507$

  $B_2 \leftarrow 30+1500 \mod 5209 = 1530$

  $B_3 \leftarrow 2600+1209 \mod 5209 = 3809$

  Now bidder divides his secret code into 3 additive shares

  $B_1$: $1000+1000+507=2507$

  $B_2$: $530+500+500=1530$

  $B_3$: $1500+809+1500=3809$

  Now each bidder computes his additive share

  $B_1$: $\sigma_1 = 1000+530+1500 = 3030$

  $B_2$: $\sigma_2 = 1000+500+809 = 2309$

  $B_3$: $\sigma_3 = 507+500+1500 = 2507$

Each bidder sends his additive share to the seller and then the seller computes his knapsack problem

$$\sigma_8 = 3030+2309+2507 = 7846 \mod 5209 = 2637$$

Now the seller starts solving the knapsack problem and sets of the flag.

$\sigma_{10} > c_{10} \Rightarrow f_{10} = 1, \sigma_9 - c_9 = 2637 - 2600 = 37;$

$\sigma_9 < (c_9, c_8, c_7, c_6, c_5) \Rightarrow f_9, f_8, f_7, f_6, f_5 = 0, \sigma_4 = 37;$

$\sigma_4 > c_4 \Rightarrow f_4 = 1, \sigma_4 - c_4 = 37 - 30 = 7;$

$\sigma_3 < (c_3) \Rightarrow f_3 = 0, \sigma_3 = 7;$

$\sigma_2 > c_2 \Rightarrow f_2 = 1, \sigma_2 - c_2 = 7 - 7 = 0$

$\sigma_1 < c_1 \Rightarrow f_1 = 0$

Final set f={0,1,0,1,0,0,0,0,0,1} and seller announces that the highest price as 1000$. The winning bidder announces himself by providing his secret 2600 to the seller. And second highest price can also be announced respectively using the same process.

This auction protocol provides an efficient distribution mechanism by solving the knapsack problem and providing security for the bids using additive sharing. We have already mentioned that the digital knapsack is not unconditionally secure. There may be need therefore to pack extra numbers into it and this can be done by increasing the number of participants and by introducing many dummy participants in a range that is sure not to be part of the winning bid. It is also essential to use strong random numbers in the use of the protocol [29].

**LIMITATIONS AND PROPOSED SOLUTIONS**

**Tie in Bidding.** This auction protocol is built assuming that no two bidders make same bid. However if two or more bidders select the same price, they are assigned with same values. This initially has no effect on the protocol but by the end of the protocol the seller will not be able to solve the knapsack problem correctly leaving room for errors. Therefore, the bidders must be able to detect the tie before computing the additive shares. So to solve this problem we propose to the use of time stamping of the bids made by the bidders. As a result the bidders who bid for the same price later than the once who have already placed the bid can be filtered out. This prevents two or more bidders to bid on same price.

**Network Time Protocol**. The Network Time Protocol is a time synchronization protocol used to obtain timestamps from a remote server on the internet. This protocol has four time values. We use the following equation $\frac{(T_4 - T_3)}{2}$ to determine the values of offset between the server and the client. The calculated offset value is used to estimate the time at the local clock in relation to the server's clock and client's clock.

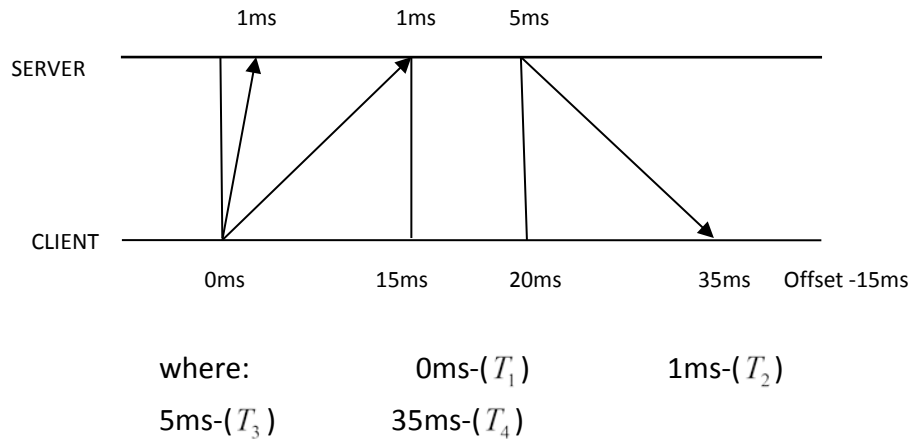

where: 0ms-($T_1$) 1ms-($T_2$)
5ms-($T_3$) 35ms-($T_4$)

Figure 4: Asymmetric propagation delay

**Attacks on Time Synchronization.** It is possible that the attacker will use the asymmetric time synchronization delays to affect the synchronization. When situations like this arise, the network time protocol uses the offset formula. Suppose the client and the server are perfectly synchronized. Let us suppose that client sends its request at 00:00:00 to the server and the server receives it at 1s. The server takes some time to process and sends the response to the client at 00:00:05. On the way back to the client the attacker delays the response and sends it at 00:00:35 on client's clock corresponding to 00:00:20 at server's clock. The attacker delays the clock by 15 seconds. The maximum error can be defined by above mentioned equation.

**Experiment.** I conducted an experiment by creating a fake auction and asked several individuals to place their bids on the prices they are willing to pay for the products I placed in the bid. I noted the timestamp depending on the bidding time of each participant. The following table is one among 5 experiments.

Whenever a bidder places his/her bid respective timestamp is recorded, this timestamp is considered when the seller encounters a tie in the bid. In the above table we can observe that there is a tie between the bidders B and A which can be solved by going through the timestamps of their bids. The timestamp of B is 2.907 whereas the timestamp of A is 2.826 which is earlier than that of B. So the tie can be broke and secret code will be assigned to A.

| Auction ID | Bidding Prices | Bidder Name | Time Stamp | Final Price |
|---|---|---|---|---|
| 1 | 120 | P | 2.654 | 120 |
| 1 | 60 | S | 2.712 | 120 |
| 1 | 80 | B | 2.907 | 120 |
| 1 | 80 | A | 2.826 | 120 |
| 1 | 100 | C | 2.900 | 120 |

Table 1: Bids with Timestamps

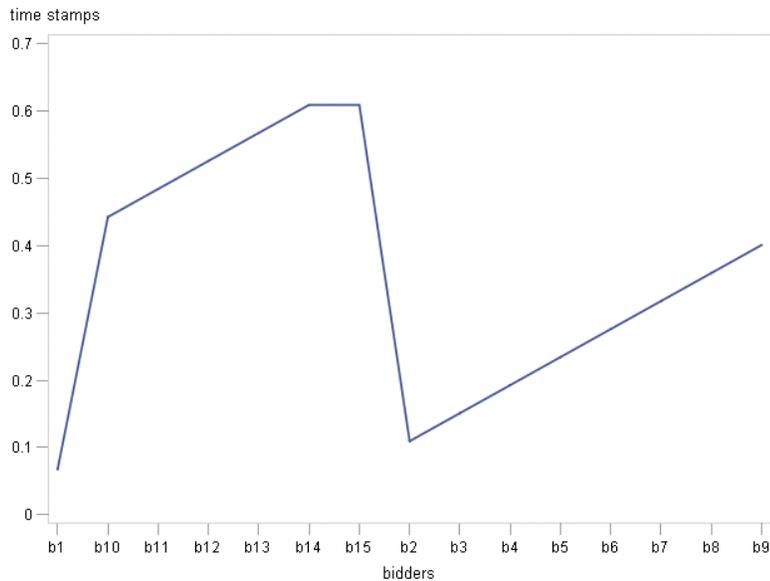

Figure 5: Timestamps for bids placed by bidders

**Selection of Price Set.** We observe that $p$ is set of prices the seller offers the bidders for their selection. The seller should make sure that the price set are selected in such a way that the count of prices is greater than the total participants in a bid. This is possible when the seller has knowledge on the number of participants in a bid prior auction. Asking the bidders to register for the auction will help the seller managing the count of the bidders.

**Registration of bidders.** All the participants of the auction are required to register for the auction. The registration may contain the Bidder Bi for its IP address, the ID of the auction he/she intends to participate and also signature of the bidder on the above information.

**Example:** Auctions in eBay uses fixed end time but allows users to bid multiple number of times. Last minute bidding is a common practice, usually called as sniping, which arises despite advise from sellers in eBay that bidders should simply submit bids that represents their maximum willingness to pay. eBay instructs bidders that they do not accept complaints on last minute bidding.

By choosing a price set before the auction gives a chance to the potential bidder to bid the price. The reason why we might see snipers on eBay is that sniping is considered the best response to a variety of strategies. But sniping in an auction with fixed deadline, in which very late bids have some probability of not being successfully placed will totally depend on the count of irrational bidders. There can be equilibria even in pure private values auctions where the bidders might have the privilege to bid late which gives higher profits to the successful bidders.

The graph of Figure 6 is plotted between the price and the bids placed in an auction. We observe that these are bids placed for a wide range of prices. Unlike eBay these auctions do not allows bidders to make multiple bids, which not only prevents sniping but also prevents the bidders from having extra privilege of winning the bid.

**Malicious Participants.** Each participant should be committed to the values he/she selects during the execution of the protocol. It may happen that the malicious participant may try to tamper with the values. This can be avoided by reconstructing the complete protocol again considering a different set of secret values. This also validates the submitted shares and also by secret keys of respective participants.

Although each bidder is allowed to place a bid, any malicious manipulation (such as tampering or manipulating another bidder's field or values) will be detected by the seller. The seller is conscious of any malicious action if a bidder has taken an error action, or it may also be possible for the seller not able to retrieve information from any field of bidder, in this case the seller can identify the tampering of the system.

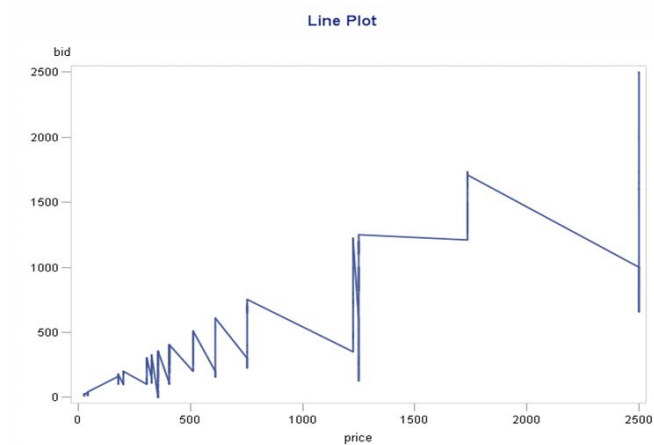

Figure 6: Bids and Prices Placed in a Bid

The steps involved in recovery of the system depend on where the error was detected.

**Disqualified bidders.** It may happen that some of the participants are disqualified due to malicious behavior. The detection of the malicious participant may affect the protocol depending on the where they are detected.

If the bidder is halted before the oblivious transfer of the secret code then the seller simply discards him from the protocol and continues the execution with the remaining bidders.

If recognized after transferring the secret code but before the additive secret sharing then the bid should be discarded.

If found after the sharing of the secret codes then the bidders are made to re-share the secret code and continue the protocol.

**CONCLUSION**

This survey analyzes the digital knapsack based secure online auction protocol. Although the digital knapsack is a cryptographically weak trapdoor function, it can be modified by increasing its size to increase its security to an extent that it is sufficient for an online auction. We describe several limitations of the protocol like detecting and solving the tie between bidders, malicious behavior of participants and also selection of price set by the seller. We propose solutions to the current limitations of the digital knapsack-based online auction protocol.